\documentclass[a4paper,11pt]{article}
\pdfoutput=1

\usepackage{jcappub} 
\usepackage{natbib}
\setcitestyle{square,comma,numbers,sort&compress}
\usepackage{enumerate}
\usepackage{tabularx}
\usepackage[left = 1.0in, top=1.0in,right=1.0in,bottom=1.0in,a4paper]{geometry}
\usepackage[dvipsnames]{xcolor}
\usepackage{comment}
\usepackage[caption=false]{subfig}
\usepackage{cancel}

\usepackage[section]{placeins}
\usepackage{listings}
\usepackage{amsbsy}
\newcolumntype{Y}{>{\centering\arraybackslash}X}

\definecolor{myRED}{rgb}{0.8, 0.25, 0.33}
\usepackage[normalem]{ulem}
\usepackage{dsfont}
\usepackage{pbox}
\usepackage{graphicx}
\usepackage{multirow}
\usepackage{tikz}
\usetikzlibrary{arrows}
\usepackage{enumitem} 
\usepackage{ulem}
\usepackage{soul}
\usepackage{lipsum} 
\usepackage{bbm}
\usepackage{amsmath}
\usepackage{amssymb}
\allowdisplaybreaks

\newcommand\scalemath[2]{\scalebox{#1}{\mbox{\ensuremath{\displaystyle #2}}}}
\usepackage{needspace}
\title{\boldmath\huge  
Neutrino masses, $\delta_\mathrm{PMNS}$, and $m_{\beta \beta}$ in SO(10)
}

\author[a]{Shaikh Saad$^\dagger$,}
\author[b]{Qaisar Shafi}

\emailAdd{$^\dagger$shaikh.saad@ijs.si}

\affiliation[a]{Jožef Stefan Institute, Jamova 39, P.\ O.\ Box 3000, SI-1001 Ljubljana, Slovenia}

\affiliation[b]{Bartol Research Institute, Department of Physics and Astronomy, University of Delaware, Newark, DE 19716, USA}

\abstract{
We explore the leptonic sector of a recently proposed supersymmetric SO(10) model with supersymmetry breaking in the 3-10 TeV range. A new ingredient in this work is the requirement that the observed baryon asymmetry is explained via non-thermal leptogenesis, which can be realized in a large class of supersymmetric hybrid inflation models including SO(10). We provide estimates for the masses of the three Standard Model neutrinos (with the lightest mass $m_1\approx 5$ meV) as well as the three right-handed neutrinos ($M_1\approx 10^9$ GeV and $M_{2,3}\approx 10^{13}$ GeV). The best fit estimate for the leptonic CP violating parameter $\delta_\mathrm{PMNS}\approx 235^\circ$, and the value of the neutrinoless double beta decay mass parameter $m_{\beta \beta}\approx 0.18$ meV.
A numerical analysis broadens the predicted range for $\delta_\mathrm{PMNS}$ ($100^\circ$--
$300^\circ$), but leaves largely intact the predictions for the six (light and heavy) neutrino masses and $m_{\beta \beta}$. Our statistical analysis, which yields the likelihood-predicted ranges of the observables, is fully consistent with JUNO’s newly released first measurement of reactor neutrino oscillations in the $\Delta m^2_{12}$–$\sin^2\theta_{12}$ plane, with JUNO improving the precision by a factor of 1.6 relative to the combination of all previous measurements. The implementation of successful non-thermal leptogenesis allows us to provide estimates for the inflaton mass ($m_\chi \approx 7\times 10^{9}$ GeV) and the reheating temperature ($T_\mathrm{RH}\approx 4\times 10^6$ GeV). 
}

\makeatletter
\gdef\@fpheader{}
\makeatother
\begin{document}
\maketitle
\flushbottom

\section{Introduction}
In a recent paper~\cite{Saad:2025cfb} we explored the phenomenology of a supersymmetric SO(10) model with the supersymmetry breaking scale in the 3-10 TeV range. Following Ref.~\cite{Babu:1998wi}, the Higgs sector of the model employs the lower dimensional SO(10) representations, with the adjoint representation being the largest. Higher dimensional operators play an essential role in this framework for explaining the observed fermion masses and mixings including the hierarchies. The best numerical fit to the data highlighted third family quasi-Yukawa unification~\cite{Gomez:2002tj,Dar:2011sj,Shafi:2023sqa}, with the MSSM parameter $\tan \beta \sim 58$. Moreover, the model predicts the three right-handed neutrino  masses, with the lightest one of order $10^9$ GeV, and the two heavier ones close to $10^{13}$ GeV.

Motivated by these considerations, in this paper we explore in particular the leptonic sector of this SO(10) model by including an important new constraint. Namely, we require that the model should also explain the observed baryon asymmetry via leptogenesis~\cite{Fukugita:1986hr}, and more specifically non-thermal leptogenesis~\cite{Lazarides:1990huy}. The reason behind the latter requirement is two fold. First, with a supersymmetry breaking scale in the 3-10 TeV range, the gravitino constraint requires the reheating temperature $T_\mathrm{RH}$ after inflation to be less than or of order $10^6-10^7$ GeV~\cite{Weinberg:1982zq,Khlopov:1984pf,Ellis:1984eq,Moroi:2005hq}. Second, as previously mentioned, the lightest right-handed neutrino mass in this SO(10) model is of order $10^9$ GeV or so, which makes it a suitable candidate for non-thermal leptogenesis following the inflationary epoch.

As it turns out an inflationary scenario with non-thermal leptogenesis can be nicely realized in a class of realistic supersymmetric hybrid inflation models~\cite{Dvali:1994ms,Rehman:2009nq,Rehman:2025fja} that we briefly discuss. We then proceed to find a best fit solution to fermion masses and mixings, taking into account the new restrictions arising from the baryon asymmetry constraint. We focus, in particular, on the leptonic sector of the model and obtain predictions for the masses of the three Standard Model (SM) as well as the three right-handed neutrinos ($N_i$, $i=1,2,3$). The best fit prediction for the leptonic Direc CP violating phase ($\delta_\mathrm{PMNS}$, PMNS here stands for Pontecorvo--Maki--Nakagawa--Sakata mixing matrix) is around $235^\circ$, and for the neutrinoless double beta decay parameter we find $m_{\beta \beta}\approx 0.18$ meV. We also perform a Markov chain Monte Carlo analysis that expands the allowed range for $\delta_\mathrm{PMNS}$, but leaves largely unchanged the other predictions. It is exciting to point out that JUNO has just released its first measurement of reactor neutrino oscillations using the first 59.1 days of data~\cite{JUNO:2025gmd}. While their measurements of $\sin^2\theta_{12}$ and $\Delta m^2_{12}$ are compatible with previous experiments, the precision is improved by a factor of 1.6 relative to the combination of all earlier measurements. Our MCMC results in the $\Delta m^2_{12}$–$\sin^2\theta_{12}$ plane are compared with JUNO’s first dataset and show full consistency. We also describe some constraints on an important parameter in the inflationary potential as well as on the reheating temperature.

The layout of the paper is as follows. In Section~\ref{sec:2}, we discuss the connection between supersymmetric hybrid inflation and non-thermal leptogenesis. Section~\ref{sec:3} describes the fermion masses and mixings in minimal SUSY SO(10), with non-renormalizable operators playing an essential role. The details of the numerical fitting procedure and the ensuing results are presented in Section~\ref{sec:4}. We summarize our results in Section~\ref{sec:5}.

\section{Supersymmetric hybrid inflation and non-thermal leptogenesis}\label{sec:2}
In this section we briefly outline a scenario based on supersymmetric hybrid inflation which, among other things, can yield the observed baryon asymmetry via non-thermal leptogenesis. The breaking of SO(10) symmetry to the left-right symmetry $SU(3)_c\times SU(2)_L\times SU(2)_R \times U(1)_{B-L}$ produces the superheavy GUT monopole~\cite{Maji:2025thf}. The primordial monopole number density can be diluted during the inflationary epoch associated with the breaking of this left-right symmetry group. The monopole can be inflated away entirely, or its number density can be reduced to observable levels, depending on the makeup of the inflationary model. Here we focus on a particularly important term in the superpotential given by
\begin{align}
    W_\mathrm{inf}= \kappa S \left(   \phi \overline \phi - M^2   \right). \label{eq:inf}
\end{align}
The `waterfall’ superfields $\phi, \overline \phi$ represent the $16_H, \overline{16}_H$ Higgs of SO(10), and $S$ denotes the SO(10) singlet inflaton field. The dimensionless parameter $\kappa$ in Eq.~\eqref{eq:inf}, it turns out, determines the inflaton mass. With a minimal superpotential and a canonical Kähler potential, the inflationary potential, including supergravity corrections, yields a scalar spectral index ($n_s$= 0.97-0.98), in excellent agreement with the current measurements~\cite{Planck:2018jri,AtacamaCosmologyTelescope:2025blo,AtacamaCosmologyTelescope:2025nti}. The $\kappa$ parameter for this minimal model is determined to be $10^{-3}$ or so, and the left-right gauge symmetry breaking scale is of order $(2-3) \times 10^{15}$ GeV. Note that in this minimal model, the waterfall fields $16_H$, $\overline{16}_H$, acquire non-zero VEVs at the end of inflation, which breaks $SU(2)_R \times U(1)_{B-L} \to U(1)_Y$, leaving supersymmetry intact. The oscillating waterfall and inflaton fields produce the right-handed neutrinos and sneutrinos from the superpotential couplings $16_i 16_j \overline{16}_H \overline{16}_H$ and $S 16_H \overline{16}_H$, where $16_i$ ($i=1,2,3$) denotes the SO(10) matter multiplet.

Following Refs.~\cite{urRehman:2006hu}, we can employ a non-minimal Kähler potential~\cite{Bastero-Gil:2006zpr}, which permits us to significantly reduce the value of $\kappa$ ($\sim 10^{-6}$), and consequently the inflaton mass. This, in turn, helps us to restrain the reheating temperature to around $10^6-10^7$ GeV, and successfully implement non-thermal leptogenesis.

To see this, it is important to note that the gauge symmetry breaking scale is essentially unchanged, namely, it remains of order $(2-3) \times 10^{15}$ GeV. With $\kappa\sim 10^{-6}$, the inflaton mass is $(2)^{1/2} \kappa M \sim (3- 4) \times 10^9$ GeV. In other words, the masses of the inflaton (and waterfall) fields are only slightly more than twice the lightest right-handed neutrino mass estimated in Ref.~\cite{Saad:2025cfb}, which helps in keeping the reheating temperature in the desired range ($10^6 - 10^7$ GeV).

Before concluding this discussion we should emphasize that with a suitable choice of the non-minimal Kähler potential~\cite{Rehman:2017gkm}, the SO(10) symmetry can be directly broken at $M_\mathrm{GUT}$ to the SM gauge group. An explicit model based  on supersymmetric SU(5) is found in Ref.~\cite{Moursy:2025ljr}. The adjoint waterfall field in this case has a non-zero VEV during inflation, which suppresses the primordial GUT monopole number density to acceptable levels. The parameter $\kappa$ in this case is of order $10^{-7}$ or so. An extension of this scenario to SO(10) lies beyond the scope of this work.

Next we turn to non-thermal leptogenesis and recall that in the instantaneous reheating approximation, the reheating temperature $T_\mathrm{RH}$ is given by~\cite{Lazarides:1996dv,Lazarides:2001zd}
\begin{align}
T_\mathrm{RH} = \left( \frac{45}{2\pi^2 g_*} \right)^{1/4} 
\frac{ ( \Gamma_\chi m_P )^{1/2} }{4}, \quad \Gamma_\chi= \frac{m_\chi}{8\pi} \frac{M_i^2}{M^2} \left( 1-4 \frac{M^2_i}{m_\chi^2} \right)^{3/2}, \label{eq:RH}
\end{align}
where, $m_P=2.4\times 10^{18}$ GeV, and we take $g_*=228.75$. As we will see, generating the correct value of the baryon asymmetry predicts that the mass of the inflaton is $m_\chi\approx 4 M_1$.

The CP-asymmetry, assuming that the inflation decays only to the lightest right-handed neutrino, is given by~\cite{Covi:1996wh}
\begin{align}
    \epsilon_1 = -\frac{1}{8\pi} \frac{1}{(h h^\dagger)_{11}} \sum_{j=2,3}
\operatorname{Im} \Bigl\{ (h h^\dagger)_{1j}^2 \Bigr\}
\left[ f_V\!\bigl(M_j^2 / M_1^2 \bigr) + f_S\!\bigl(M_j^2 / M_1^2 \bigr) \right], \label{eq:epsilon}
\end{align}
where
\begin{align}
    f_V(x) = \sqrt{x} \, \ln\!\Bigl(1 + \frac{1}{x}\Bigr), 
\qquad
f_S(x) = \frac{2\sqrt{x}}{x - 1} .
\end{align}
Here, $h$ is the  neutrino Dirac Yukawa coupling matrix ($W\supset h_{i\alpha} N_i L_\alpha H_u$) which is defined in the basis where both the charged lepton and the right-handed neutrino mass matrices are real and diagonal.   The  lepton asymmetry is given by~\cite{Senoguz:2003hc}
\begin{align}
    \frac{n_L}{s} = \frac{3}{2} \; \frac{T_\mathrm{RH}}{m_\chi} \; \sum_i \epsilon_i \; \mathrm{Br}(\chi \to N_i N_i) .\label{eq:etaB}
\end{align}
With $m_\chi < 2 M_{2,3}$, $\mathrm{Br}(\chi \to N_1 N_1) =1$. To replicate the observed~\cite{Planck:2018vyg} baryon asymmetry of the universe, one therefore requires
\begin{align}
&\eta_B\equiv \frac{n_B}{n_\gamma}=6.1\times 10^{-10},    
\quad
\frac{n_B}{s} \simeq \frac{\eta}{7.04},
\quad
\frac{n_L}{s}=-\frac{79}{28} \frac{n_B}{s} = -2.44\times 10^{-10}.
\end{align}

\section{Fermion Masses in Supersymmetric SO(10)}\label{sec:3}
To trace the origin of fermion masses within the SO(10) framework, one examines the following tensor product: 
\begin{equation}
16\times 16 = 10_\mathrm{s}+ 120_\mathrm{a} + 126_s. \label{eq:bilinear}
\end{equation}
Here the subscripts s and a stand for symmetric and antisymmetric components (in family
space). It turns out that in SUSY SO(10), larger representations such as \(120\) and \(126\) lead to non-perturbative gauge couplings just above the GUT scale. Therefore, in this work following~\cite{Saad:2025cfb}, we restrict ourselves to lower-dimensional representations. In particular, we restrict ourselves to representations with dimensionality no larger than the adjoint representation. The minimal Higgs sector therefore consists of  $10_H$,  $45_H$,   and a pair of $16_H+\overline{16}_H$. The SO(10) symmetry can be broken directly to the MSSM, or via the chain :
\begin{align}
SO(10)
&\xrightarrow[45_H]{a} 
SU(3)_c\times SU(2)_{L} \times SU(2)_{R}\times U(1)_{B-L} \label{eq:SSB:01}
\\
&  \xrightarrow[16_H+ \overline{16}_H]{c, \overline{c}}
SU(3)_c\times SU(2)_L\times U(1)_Y \label{eq:SSB:02}
\\
&  \xrightarrow[10_H, 16_H+ \overline{16}_H]{}
SU(3)_c\times  U(1)_\mathrm{em}.  \label{eq:SSB:03}
\end{align} 
In this chain, the first stage of the symmetry breaking can be obtained with the following VEV structure for the adjoint field, $\langle 45_H\rangle= i\tau_2\otimes \mathrm{diag}(a,a,a,0,0)$. The intermediate symmetry is spontaneously broken close to $M_\mathrm{GUT}$, it turns out,  by the spinorial representations with VEV $\langle 16_H\rangle = c,  \langle \overline{16}_H\rangle= \overline c$. We assume a hierarchical structure in the magnitude of the VEVs, $a \gtrsim c, \overline{c} \gg v_\mathrm{EW}$.

\begin{figure}[t!]
\centering
\includegraphics[width=0.5\textwidth]{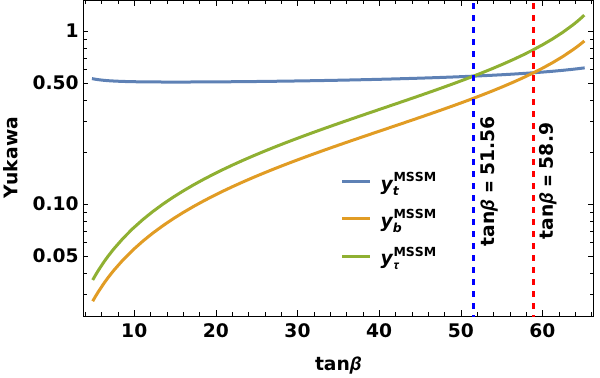}
\caption{Third generation Yukawa couplings at  $M_\mathrm{GUT}=2\times 10^{16}$ GeV, as a function of $\tan\beta$ for $M_\mathrm{SUSY}=3$ TeV, ignoring threshold corrections.  $t\!-\!b$ Yukawa unification occurs at $\tan\beta \sim 58.9$ (red dashed vertical line), whereas $t\!-\!\tau$ unification corresponds to $\tan\beta \sim 51.56$  (blue dashed vertical line).  }\label{fig:Yukawas}
\end{figure}

As for the fermion masses, we first focus on the third generation fermions, with the renormalizable Yukawa interaction 
\begin{align}
W_{\text{Yukawa}}\supset  h_{33}  16_316_310_H. \label{eq:01}
\end{align}
At $M_\mathrm{GUT}$, one obtains~\cite{Ananthanarayan:1991xp}
\begin{align}
y^\mathrm{MSSM}_t(M_\mathrm{GUT})=     y^\mathrm{MSSM}_b(M_\mathrm{GUT}) = y^\mathrm{MSSM}_\tau(M_\mathrm{GUT}). 
\end{align}
However, with (multi-)TeV scale SUSY breaking and neglecting SUSY threshold corrections, the experimentally measured values do not yield exact $t\!-\!b\!-\!\tau$ unification at the GUT scale. This is shown in  Fig.~\ref{fig:Yukawas}, with data taken from Ref.~\cite{Antusch:2013jca} for $M_\mathrm{SUSY}=3$ TeV and $M_\mathrm{GUT}=2\times 10^{16}$ GeV. The figure shows that $t\!-\!b$ Yukawa unification occurs at $\tan\beta \sim 58.9$ (red dashed vertical line), and $t\!-\!\tau$ Yukawa unification corresponds to $\tan\beta \sim 51.56$  (blue dashed vertical line). For the former (latter) value of $\tan\beta$, one finds $y_\tau^\mathrm{MSSM}/y_b^\mathrm{MSSM} \approx 1.36$ ($y_t^\mathrm{MSSM}/y_b^\mathrm{MSSM} \approx 1.34$).

As proposed in Ref.~\cite{Saad:2025cfb}, in addition to the dimension-four operator given in Eq.~\eqref{eq:01}, one introduces the following dimension-six operator, which yields distinct contributions to the bottom and tau masses:
\begin{align}
W_{\text{Yukawa}}\supset  \frac{s_{33}}{\Lambda^2}  16_316_310_H45_H^2,   \label{eq:03}
\end{align}
Through this operator, a quasi-Yukawa unification of the third-generation couplings consistent with the above data can be achieved, namely, $y_t^\mathrm{MSSM} \approx y_b^\mathrm{MSSM} \approx 0.73\, y_\tau^\mathrm{MSSM}$.

Moreover, to reproduce all of the observed charged fermion masses and mixings within the minimal SUSY $\mathrm{SO}(10)$ framework, the minimal set of operators required, including   the two discussed above, is given by~\cite{Saad:2025cfb}  
\begin{align}
   \scalemath{0.95}{ W_{\text{Yukawa}} = 
Y_{10}^{ij} \, 16_i \, 16_j \, 10_H + 
\frac{1}{\Lambda} Y_{45}^{ij} \, 16_i \, 16_j \, 10_H \, 45_H + 
\frac{1}{\Lambda} Y_{16}^{ij} \, 16_i \, 16_j \, 16_H \, 16_H 
+ \frac{1}{\Lambda^2} Y_{s}^{ij} \, 16_i \, 16_j \, 10_H \, 45_H^2,}
\end{align}
such that
\begin{align}
&Y_{10}^\mathrm{sym}= \begin{pmatrix}
h_{11}&h_{12}&0\\
h_{12}&0&h_{23}\\
0&h_{23}&h_{33}
\end{pmatrix},
Y_{45}^\mathrm{anti}= \begin{pmatrix}
0&a_{12}&0\\
-a_{12}&0&a_{23}\\
0&-a_{23}&0
\end{pmatrix},
Y_{16}^\mathrm{sym}= \begin{pmatrix}
0&g_{12}&0\\
g_{12}&0&g_{23}\\
0&g_{23}&0
\end{pmatrix},
Y_{s}^\mathrm{sym}= \begin{pmatrix}
0&0&0\\
0&0&0\\
0&0&s_{33}
\end{pmatrix}.
\end{align}
The fermion mass matrices, in the basis $f M_f f^c$, are given by:
\begin{align}
&\frac{M_U}{v_u}= Y_U^\mathrm{MSSM}= \bigg\{Y_{10}^\mathrm{sym}   +  \xi_{45} 
 \; Y_{45}^\mathrm{anti}
+
\xi_{45}^2
\; Y_{s}^\mathrm{sym}
\bigg\}, \label{eq:up}
\\
&\frac{M^D_\nu}{v_u}=  \left(Y^D_\nu\right)^\mathrm{MSSM}=  \bigg\{Y_{10}^\mathrm{sym}  -3  \xi_{45}  \; Y_{45}^\mathrm{anti}
+
9\xi_{45}^2
\; Y_{s}^\mathrm{sym}
\bigg\},
\\
&\frac{M_D}{v_d}=  Y_D^\mathrm{MSSM}=  \cos\gamma \bigg\{Y_{10}^\mathrm{sym}  +  \xi_{45} \;  Y_{45}^\mathrm{anti}
+
\xi_{45}^2
\;Y_{s}^\mathrm{sym} 
- \xi_{16} \tan\gamma \;Y_{16}^\mathrm{sym}
\bigg\},
\\
&\frac{M_E}{v_d}= Y_E^\mathrm{MSSM}=  \cos\gamma \bigg\{Y_{10}^\mathrm{sym}    -3 \xi_{45} \; Y_{45}^\mathrm{anti}
+
9
 \xi_{45}^2
\;\overline Y_{s}^\mathrm{sym}
- \xi_{16} \tan\gamma  \;Y_{16}^\mathrm{sym}
\bigg\}.  \label{eq:lepton}
\end{align}
Here,  we have defined
$\xi_{45}= a/\Lambda$  and $\xi_{16}=  c/\Lambda$ (recall, $|c|=|\overline c|$). To ensure the validity of the effective theory, we restrict ourselves to the case $\xi_{45,16} \leq 0.2$. Since there are two down-type Higgs doublets, one originating from  $10_H$ and the other from $16_H$, we define:
\begin{align}
&H_u=10_u,\quad H_d=\cos\gamma 10_d + \sin\gamma 16_d,
\\
&\tan\beta= \frac{v_u}{v_d},\quad 
\tan\gamma= \frac{\langle 10_d\rangle}{\langle 16_d\rangle},
\end{align}
where
\begin{align}
v^{10}_u=v\sin\beta, \quad   
v^{10}_d=v\cos\beta\cos\gamma, \quad   
v^{16}_d=v\cos\beta\sin\gamma,
\quad v=174.104\mathrm{GeV},
\end{align}
and as usual, $\langle H_u\rangle= v_u=v\sin\beta$, $\langle H_d\rangle=v_d=v\cos\beta$.

The mass matrices can be rewritten in a more convenient form as follows: 
\begin{align}
&M_U = 
\begin{pmatrix}
r_1 & \hspace{5pt} \epsilon' +r_2& 0 \\
-\epsilon' +r_2& 0 & \hspace{7pt} \epsilon + \sigma \\
0 & \hspace{5pt} -\epsilon + \sigma & \hspace{5pt} 1+\zeta
\end{pmatrix} m_U, \quad
M_D = 
\begin{pmatrix}
r_1 & {}\epsilon' + \eta^{\prime\prime}& 0 \\
-\epsilon' + \eta^{\prime\prime}& 0 & {}\epsilon + \eta \\
0 & \hspace{5pt} -\epsilon + \eta & \hspace{9pt} 1+\zeta
\end{pmatrix} m_D, \label{eq:mass01}
\\&
M^D_\nu = 
\begin{pmatrix}
r_1 & -3\epsilon' +r_2& 0 \\
3\epsilon' +r_2& 0 & -3\epsilon + \sigma \\
0 & 3\epsilon + \sigma & 1+9\zeta
\end{pmatrix} m_U, \quad
M_E = 
\begin{pmatrix}
r_1 & -3\epsilon' + \eta^{\prime\prime}& 0 \\
3\epsilon' + \eta^{\prime\prime}& 0 & -3\epsilon + \eta \\
0 & 3\epsilon + \eta & 1+9\zeta
\end{pmatrix} m_D. \label{eq:mass02}
\end{align}
In writing these, we have defined the following quantities: 
\begin{align}
&m_U= v\sin\beta\; h_{33},\quad m_D= v\cos\beta \cos\gamma\;  h_{33}, \quad \zeta=\xi_{45}^2\;  \frac{ s_{33}}{ h_{33}},
\quad
\epsilon= \frac{ a_{23}}{ h_{33}}\xi_{45},\quad
\sigma= \frac{ h_{23}}{ h_{33}},
\\&
\eta= \sigma -  \frac{ g_{23}}{h_{33}}\xi_{16}\tan\gamma,
\quad 
\epsilon^\prime= \frac{ a_{12}}{ h_{33}}\xi_{45},\quad
\eta^\prime= - \frac{ g_{12}}{ h_{33}}\xi_{16}\tan\gamma, 
\quad 
r_1=\frac{ h_{11}}{ h_{33}}, \quad r_2=\frac{ h_{12}}{ h_{33}}, \quad \eta^{\prime\prime}=\eta' +r_2.
\end{align}
Note that due to the hierarchical structure of the charged fermion masses, to a very good approximation, one obtains
\begin{align}
    \frac{m_t}{m_b}\approx \frac{m_U}{m_D} = \frac{\tan\beta}{\cos\gamma}, \quad  \mathrm{and}  \quad  \frac{m_\tau}{m_b}\approx \frac{1+9\zeta}{1+\zeta}\approx 1+9\zeta. 
\end{align}

Finally, the right-handed Majorana mass matrix arises from 
\begin{align}
  W_{\text{Yukawa}} \supset \frac{1}{\Lambda} Y^{ij}_{\nu^c} 16_i16_j \overline{16}_H\overline{16}_H .
\end{align}
The SM neutrinos receive their masses through the seesaw mechanism~\cite{Minkowski:1977sc,Gell-Mann:1979vob,Yanagida:1979as,Schechter:1980gr,Glashow:1979nm,Mohapatra:1979ia} 
\begin{align}
\mathcal{M}_\nu=- M^D_\nu   (M_\nu^R)^{-1}    \left(M^D_\nu\right)^T , \quad  M_\nu^R=\frac{\overline c^2}{\Lambda} Y_{\nu^c}.  \label{eq:mass03}
\end{align}
Following Ref~\cite{Babu:1998wi}, we take an economical ansatz for the right-handed Majorana mass matrix
\begin{align}
    &M_\nu^R=M_R \begin{pmatrix}
x&0&z\\
0&0&y\\
z&y&1
\end{pmatrix},
\quad  
x=\frac{\left(Y_{\nu^c}\right)_{11}}{\left(Y_{\nu^c}\right)_{33}}, \quad 
z=\frac{\left(Y_{\nu^c}\right)_{12}}{\left(Y_{\nu^c}\right)_{33}}, \quad
y=\frac{\left(Y_{\nu^c}\right)_{23}}{\left(Y_{\nu^c}\right)_{33}}, \quad
M_R=\frac{\overline c^2}{\Lambda} \left(Y_{\nu^c}\right)_{33}. \label{xyz}
\end{align}
We will have more to say about the parameters $x, y, z$ below.

\section{Numerical analysis and results}\label{sec:4}
\textbf{Fit procedure:} 
For the numerical analysis we perform a $\chi^2$-analysis which minimizes the following function:
\begin{align}
\chi^2= \sum_i \mathrm{pull}_i^2,\;\;\; \;\;\mathrm{pull}_i= \frac{T_i-O_i}{\sigma_i}.    
\end{align} 
Here, \(O_i\), \(\sigma_i\), and \(T_i\) denote, respectively, the experimentally measured value, the associated uncertainty, and the theoretical prediction for the observable labeled by \(i\). The summation over \(i\) includes all observables, namely, the six quark masses, the four CKM mixing parameters (including the CP-violating Dirac phase), the three charged-lepton masses, the two neutrino mass-squared differences, and the three leptonic mixing angles---18 observables in total.  In addition to the fermion masses and mixing parameters, we also include a 19th observable, namely, the baryon asymmetry parameter $\eta$, for which we allow an uncertainty of $10\%$. Note that the CP-violating Dirac phase in the PMNS   mixing matrix has not yet been measured experimentally, and thus its value is not included in the $\chi^2$-function.

For the experimental values of the charged fermion masses and mixings at the GUT scale, we use the data provided in Ref.~\cite{Antusch:2013jca} for a scenario with zero threshold corrections. Assuming \clearpage  \noindent vanishing threshold corrections\footnote{As will be discussed later, our model admits solutions with $\tan\beta \sim 50$. In such scenarios, relatively small (few percent) or negligible threshold corrections can be obtained through an appropriate hierarchical choice of MSSM parameters. For illustration, consider the correction to the bottom-quark mass~\cite{Hall:1993gn,Elor:2012ig}:
\begin{align}
m_b\simeq v_d y_b \bigg\{  1+ 0.35 \left( \frac{2 \mu m_{\widetilde g}}{m^2_0}  + \frac{\mu A_t}{m^2_0}\right) \bigg\}.     
\end{align}
Fixing the SUSY scale at $m_0 = 3\text{TeV}$ and taking the gluino mass $m_{\widetilde g} \sim 2\text{TeV}$, consistent with current LHC bounds, the threshold corrections remain small provided $\mu, A_t \lesssim 100,\text{GeV}$~\cite{Hall:1993gn}.
}, the matching conditions between the SM and MSSM Yukawa couplings at the SUSY scale are given by~\cite{Antusch:2013jca}
\begin{align}
    Y_U^\textrm{SM}=Y_U^\textrm{MSSM}\sin\beta\,,
    \quad
    Y_D^\textrm{SM}=Y_D^\textrm{MSSM}\cos\beta\,,
    \quad
    Y_E^\textrm{SM}=Y_E^\textrm{MSSM}\cos\beta\,.  \label{eq:matching}
\end{align}
Using \texttt{REAP}~\cite{Antusch:2005gp}, which implements two-loop RGEs, these quantities are evolved from the SUSY scale, $M_S=3$ TeV, to the GUT scale, $M_\mathrm{GUT}=2\times 10^{16}$ GeV, and the corresponding values are obtained as functions of $\tan\beta$~\cite{Antusch:2013jca}. These values of charged fermion masses and mixings are taken as input parameters for our fit. The values of the  neutrino oscillation parameters used in our fit are taken from Ref.~\cite{Esteban:2024eli,NUFIT}. For the neutrino observables, during the fitting procedure we allow for up to a 5\% error to ensure numerical stability. Nevertheless, we obtain excellent agreement with the experimentally measured values.

Inspecting the structure of the charged fermion mass matrices in Eqs.~\eqref{eq:mass01}--\eqref{eq:mass02}, one finds that they contain ten parameters: \(m_{U,D}\), \(\zeta\), \(\epsilon^{(\prime)}\), \(\eta^{(\prime)}\), \(\sigma\), and \(r_{1,2}\). The Majorana neutrino mass matrix  involves four additional parameters---a mass scale \(M_R\) and three  ratios \(x\), \(y\), and \(z\) of the Yukawa couplings (see Eq.~\eqref{xyz}). For the purpose of fitting fermion masses and mixing angles, all parameters can initially be taken as real, yielding a total of 14 real parameters (magnitudes) to reproduce 18 observables (including the magnitude of $\eta_B$). This shows that the system is highly constrained and thus predictive. Since the CKM matrix exhibits a CP-violating phase, this feature is incorporated by allowing two complex parameters in the \(1\!-\!2\) sector. Specifically, the Yukawa couplings \(h_{21}\) and \(a_{21}\) are taken to be complex, corresponding to complex \(r_2\) and \(\epsilon'\), respectively. Consequently, the Dirac neutrino mass matrix acquires complex entries, and a consistent fit to the neutrino oscillation data further requires the ratios \(x\), \(y\), and \(z\) to be complex. All of these phases play an important role in determining the baryon asymmetry parameter as well as predicting the Dirac CP-violating phase in the PMNS matrix.  Taking these phases into account, the model contains 14 magnitudes and 5 phases to fit a total of 19 observables (18 magnitudes and one CP-violating phase).

\begin{table}[t!]
\centering
\footnotesize
\resizebox{0.7\textwidth}{!}{
\begin{tabular}{|c|c|c|c|}
\hline
\textbf{Observables} & \textbf{Expt. Values} & \textbf{  Fitted Values} &\textbf{Pulls} \\
\hline\hline

$y_u\sin\beta/10^{-6}$        & 2.7612 & 2.7884 & 0.031\\  
$y_c\sin\beta/10^{-3}$        & 1.4319 & 1.4523 & 0.285\\ 
$y_t\sin\beta$                & 0.5536 & 0.5516 & -0.072\\ \hline

$y_d\cos\beta/10^{-5}$        & 0.6288 & 0.4116 & -1.727\\ 
$y_s\cos\beta/10^{-4}$        & 1.2447 & 1.1991 & -0.733\\ 
$y_b\cos\beta/10^{-2}$        & 0.8248 & 0.8344 & 0.232\\ \hline

$y_e\cos\beta/10^{-6}$        & 2.5604 & 2.5642 & 0.027\\ 
$y_\mu\cos\beta/10^{-4}$      & 5.4082 & 5.4789 & 0.262\\ 
$y_\tau\cos\beta/10^{-2}$     & 1.1101 & 1.1298 & 0.355\\ \hline

$\theta_{12}^{CKM}$           & 0.2273 & 0.2340 & 0.587\\ 
$\theta_{23}^{CKM}/10^{-2}$   & 3.7182 & 3.8806 & 0.874\\ 
$\theta_{13}^{CKM}/10^{-3}$   & 3.2356 & 3.2424 & 0.041\\ 
$\delta_{CKM}^{c}$            & 1.2080 & 1.2363 & 0.469\\
\hline\hline

$\Delta m^2_{21} \text{ (eV}^2)/10^{-5}$ & 7.4900 & 7.4938 & 0.010\\ 
$\Delta m^2_{31} \text{ (eV}^2)/10^{-3}$ & 2.5345 & 2.5307 & -0.029\\ \hline

$\sin^2 \theta_{12}$          & 0.3075 & 0.3081 & 0.037\\ 
$\sin^2 \theta_{23}$          & 0.5596 & 0.5561 & -0.124\\ 
$\sin^2 \theta_{13}$          & 0.0219 & 0.0219 & -0.025\\

$\eta_B/10^{-10}$ & $6.1$ &  6.03 &    -0.11 \\
\hline\hline

$\chi^2$& - & - & 5.21  \\
\hline
\end{tabular}
}
\caption{Input values of the observables in the charged fermion sector  at  $M_\mathrm{GUT}=2\times 10^{16}$ GeV for a  SUSY breaking scale $M_S=3$ TeV,  taken from Ref.~\cite{Antusch:2013jca} (corresponding to our best fit  $\tan\beta=53.3226$). The experimental values of the observables in the neutrino sector are taken from Ref.~\cite{NUFIT}.    The uncertainty associated with the baryon asymmetry parameter $\eta_B$ is taken to be 10\%.}  
\label{result}
\end{table}

\textbf{Obtaining the best fit:} 
For the numerical analysis, we fix the inflation scale\footnote{Although the inflation scale can approach the GUT scale, our numerical analysis shows this  tends to increase the total $\chi^2$ value.} to $M = 3 \times 10^{15}~\mathrm{GeV}$, while the cutoff scale is taken to be $\Lambda = 10^{17}~\mathrm{GeV}$, as expected in this theory (see, e.g., Ref.~\cite{Saad:2025cfb}). In this case, $\xi_{45}=M_\mathrm{GUT}/\Lambda=0.2$ and $\xi_{16}=M/\Lambda=0.03$.  
After performing an extensive numerical search of the parameter space we obtain the best fit corresponding to a total\footnote{In this work, we slightly improve the total $\chi^2$ compared to our previous analysis~\cite{Saad:2025cfb}. In the earlier study, the charged fermion sector was fitted separately, followed by a fit to the neutrino sector. In contrast, here we perform a combined fit to all sectors simultaneously.} $\chi^2=5.2$ with 
\begin{align}
&m_U= 89.7817\;\mathrm{GeV},\quad m_D= 1.34246\;\mathrm{GeV}, \quad M_R= 6.21437\times 10^{12}\mathrm{GeV},
\\
&\epsilon= -0.122438,\quad
\epsilon^\prime= 8.46077\times 10^{-4}e^{1.77311 i},\quad \sigma= 0.132364,
\\
&\eta=0.176395,\quad \eta^\prime=-3.29535\times 10^{-3}, \quad \zeta= 3.8777\times 10^{-2},
\\
&r_1= -1.29647\times 10^{-4},\quad r_2=6.6853\times 10^{-4}   e^{-1.20929 i},
\\
&x=2.93974\times 10^{-4} e^{2.80275 i},\quad z=6.52862\times 10^{-2} e^{1.05198 i},\quad y=1.2258 e^{-0.74676 i}.
\end{align}
The corresponding best fit values of the observables are listed in Table~\ref{result}, and the best fit predictions of some of the observables are summarized in Table~\ref{tab:fit}.

\begin{table}[t!]
\centering
\renewcommand{\arraystretch}{1}
\begin{tabular}{c c c}
\hline
Observable & Best fit predictions & $3\sigma$ HPD interval\\
\hline
$m_1 (meV)$ & 5.12 & $2.95-7.14$ \\ 
$m_2 (meV)$ & 10.05 & $8.79-11.30$\\
$m_3 (meV)$ & 50.56 & $46.60-54.26$\\  \hline

$m_{\beta\beta} (meV)$ & 0.18 & $0.14-0.21$  \\  \hline

$\delta_\mathrm{PMNS}$ (deg)        & $234.96$  & $90.37-308.97$ \\ \hline

$M_1$ (GeV)  &  $1.82\times 10^{9}$ & $(1.32-3.32)\times 10^{9}$ \\ 
$M_2$ (GeV)  &  $7.42\times 10^{12}$ & $(5.76-10.62)\times 10^{12}$\\ 
$M_3$ (GeV)  &  $7.83\times 10^{12}$ & $(6.09-11.24)\times 10^{12}$\\ \hline

$m_\chi$ (GeV)  &  $7.29\times 10^{9}$ & $(4.01-48.98)\times 10^{9}$\\ 
$T_\mathrm{RH}$  (GeV) & $4.08\times 10^6$ & $(1.41-20.08)\times 10^6$\\
$\kappa$ & $1.72\times 10^{-6}$ & $(0.94-17.84)\times 10^{-6}$\\ \hline

\end{tabular}
\caption{Best fit predictions for the six neutrino masses ($m_i$ and $M_i$),  $\delta_\mathrm{PMNS}$, and $m_{\beta\beta}$. Also listed are values for the inflaton mass, the reheating temperature, and the inflationary potential parameter $\kappa$. Alongside the best-fit prediction, the last column presents the $3\sigma$ highest posterior density (HPD) interval derived from the MCMC analysis.}
\label{tab:fit}
\end{table}

For the best fit solution, with $\tan\beta=53.3226$ and $\cos\gamma=0.797306$, we obtain
\begin{align}
\frac{m_t}{m_b}\approx \frac{m_U}{m_D} = \frac{\tan\beta}{\cos\gamma}=66.8784, \quad  \mathrm{and}  \quad  \frac{m_\tau}{m_b}\approx \frac{1+9\zeta}{1+\zeta}\approx 1+9\zeta=1.34899. 
\end{align}
Furthermore, using  Eqs.~\eqref{eq:up}-\eqref{eq:lepton} and the matching conditions in Eq.~\eqref{eq:matching}, we find
\begin{align} 
&y_t^\mathrm{MSSM}/y_b^\mathrm{MSSM}\approx (\cos\gamma)^{-1}=1.254, \\ 
&y_\tau^\mathrm{MSSM}/y_b^\mathrm{MSSM}\approx 1+9\zeta=1.348, \label{second} \\
&y_t^\mathrm{MSSM}/y_\tau^\mathrm{MSSM}\approx (\cos\gamma (1+9\zeta))^{-1}=0.929.    
\end{align}
To emphasize, Eq.~\eqref{second} highlights the crucial role played by the dimension six operator in Eq.~\eqref{eq:03} in splitting the bottom and tau Yukawa couplings. The results above show that the third family quasi-Yukawa unification relations are slightly modified in this SO(10) model.

Although for low value, say $\tan\beta= 10$,  a consistent fit to the fermion masses and mixings is possible~\cite{Saad:2025cfb}, a simultaneous fit that also reproduces the correct baryon asymmetry seems difficult to achieve. Therefore, $\tan\beta \sim 53$ appears to be the optimal value for achieving a simultaneous fit.

\begin{figure}[b!]
\centering
\includegraphics[width=0.48\textwidth]{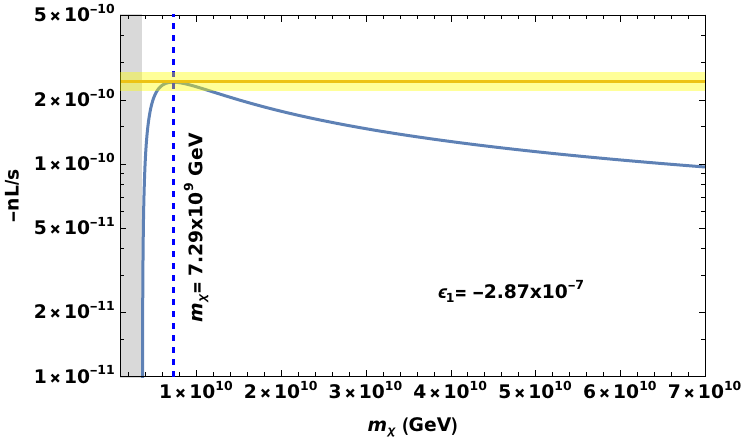}
\includegraphics[width=0.43\textwidth]{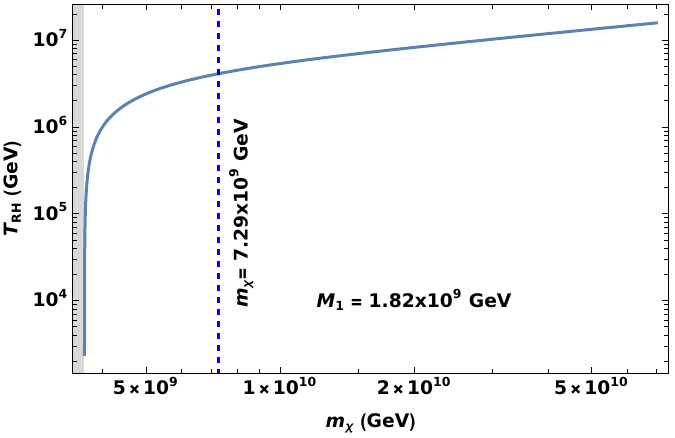}
\caption{The left panel shows, for the best fit (that has $\epsilon_1=-2.87\times 10^{-7}$), the  inflaton mass $m_\chi$ consistent with correctly reproducing the baryon asymmetry. The corresponding reheating temperature $T_\mathrm{RH}$ is shown in the right panel as a function of $m_\chi$ (where $T_{\mathrm{RH}} \approx 4 \times 10^{6}\,\mathrm{GeV}$ is compatible with the observed baryon asymmetry). In the gray shaded region the inflaton decay into $N_1$ is kinematically forbidden. The horizontal yellow band shows a 10\% variation of the baryon asymmetry parameter around its central value. }\label{fig:mchi}
\end{figure}
\textbf{Fit results and parameter space exploration:} 
As mentioned above, the fitting procedure also includes the baryon asymmetry parameter. Remarkably, our model successfully reproduces the observed baryon asymmetry of the Universe. Furthermore, in achieving this, the model predicts the inflaton mass , namely,   $m_\chi \approx 4 M_1$, as well as the reheating temperature $T_\mathrm{RH} \sim 10^6~\mathrm{GeV}$  (see Table~\ref{tab:fit}). This result arises from the interplay between the fermion mass fit and the requirement of generating the correct baryon asymmetry with only a limited number of parameters. To illustrate this correlation, we depict the relationship between the baryon asymmetry parameter and the inflaton mass in Fig.~\ref{fig:mchi}.

\begin{figure}[t!]
\centering
\includegraphics[width=0.45\textwidth]{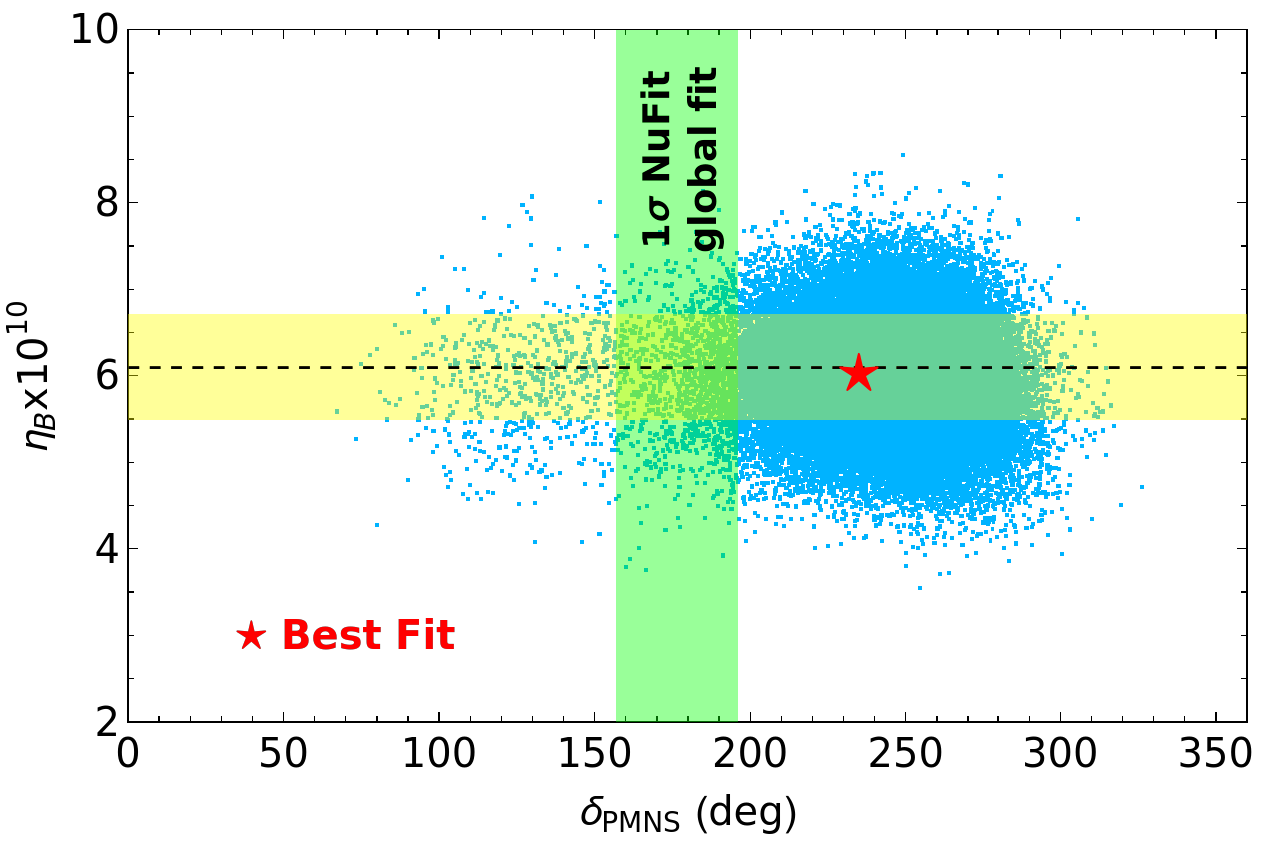}
\includegraphics[width=0.46\textwidth]{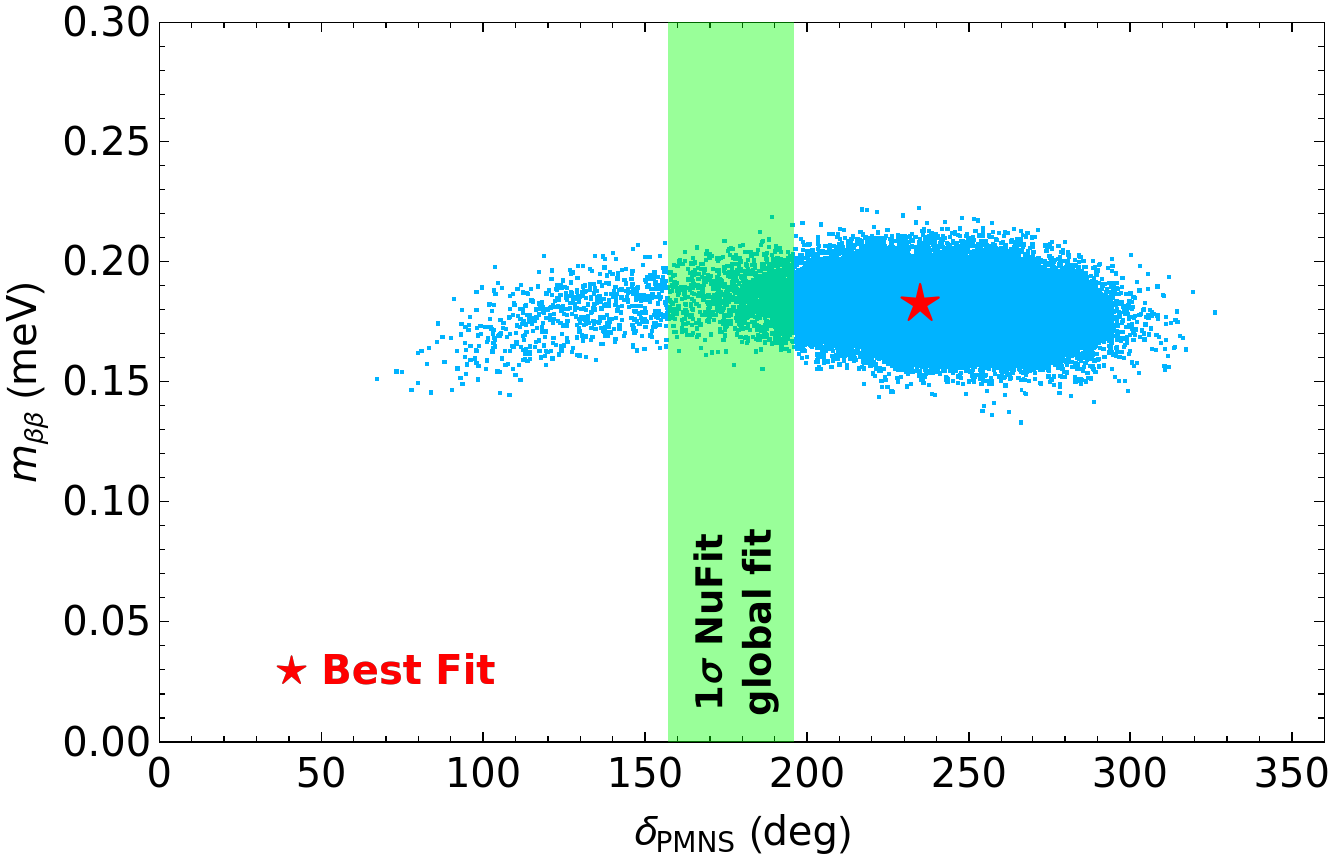}
\caption{MCMC result: The likelihood range of $\delta_\mathrm{PMNS}\in (100,300)^\circ$--compatible with fermion masses and baryon asymmetry with the correct sign and magnitude. Although we have shown the $1\sigma$ range of $\delta_\mathrm{PMNS}$ from NuFit, the $3\sigma$ range spans a much wider interval of $96^\circ - 422^\circ$~\cite{NUFIT}. The red star corresponds to the best fit solution ($\delta_\mathrm{PMNS}=235^\circ$). 
 } \label{fig:001}
\end{figure}
In the left panel of this figure, the lepton asymmetry is shown as a function of the inflaton mass for the best-fit solution. As seen from Eq.~\eqref{eq:etaB}, the lepton asymmetry is evaluated by the inflaton mass, the right-handed neutrino mass $M_1$, and the CP-asymmetry parameter $\epsilon_1$ Eq.~\eqref{eq:epsilon}. Since our model contains only a limited number of free parameters, a key parameter, $\epsilon_1$, which is determined by the fermion mass fit\footnote{
Since the right-handed neutrino masses $M_i$ are completely fixed in this framework,   $\epsilon_1$ is determined exclusively by the restricted structure of the Yukawa couplings.
}, cannot be made arbitrarily large in absolute magnitude. As a result, once the constraint on the baryon asymmetry is imposed, its magnitude (with the correct sign) is essentially determined by minimality and is of order $\mathcal{O}(10^{-7})$. It should be emphasized that the absolute value of $\epsilon_1$ can be smaller than $\mathcal{O}(10^{-7})$ if the baryon asymmetry constraint is not imposed. This will be demonstrated explicitly using the MCMC results.

Given this value of $\epsilon_1$ and the predicted mass $M_1 \approx 10^9\,\mathrm{GeV}$, achieving a successful baryon asymmetry consistent with observations constrains the inflaton mass to be $m_\chi \sim 7 \times 10^{9}\,\mathrm{GeV} \approx 4 M_1$. In short, reproducing the correct magnitude and sign of the baryon asymmetry leads to a prediction for the inflaton mass, implying that higher inflaton masses are disfavored in this framework. The yellow horizontal band in Fig.~\ref{fig:mchi} represents a 10\% variation of the asymmetry around its central value, while the gray vertical band indicates the region where inflaton decay into $N_1$ is kinematically forbidden. Moreover, the best-fit value of the inflaton mass is indicated by the vertical blue dashed line. Subsequently, the reheating temperature $T_\mathrm{RH}$, which depends on $M_1$ and $m_\chi$ as given in Eq.~\eqref{eq:RH}, is also predicted. The resulting prediction for the reheating temperature is shown in the right panel of Fig.~\ref{fig:mchi}.

\begin{figure}[t!]
\centering
\includegraphics[width=0.46\textwidth]{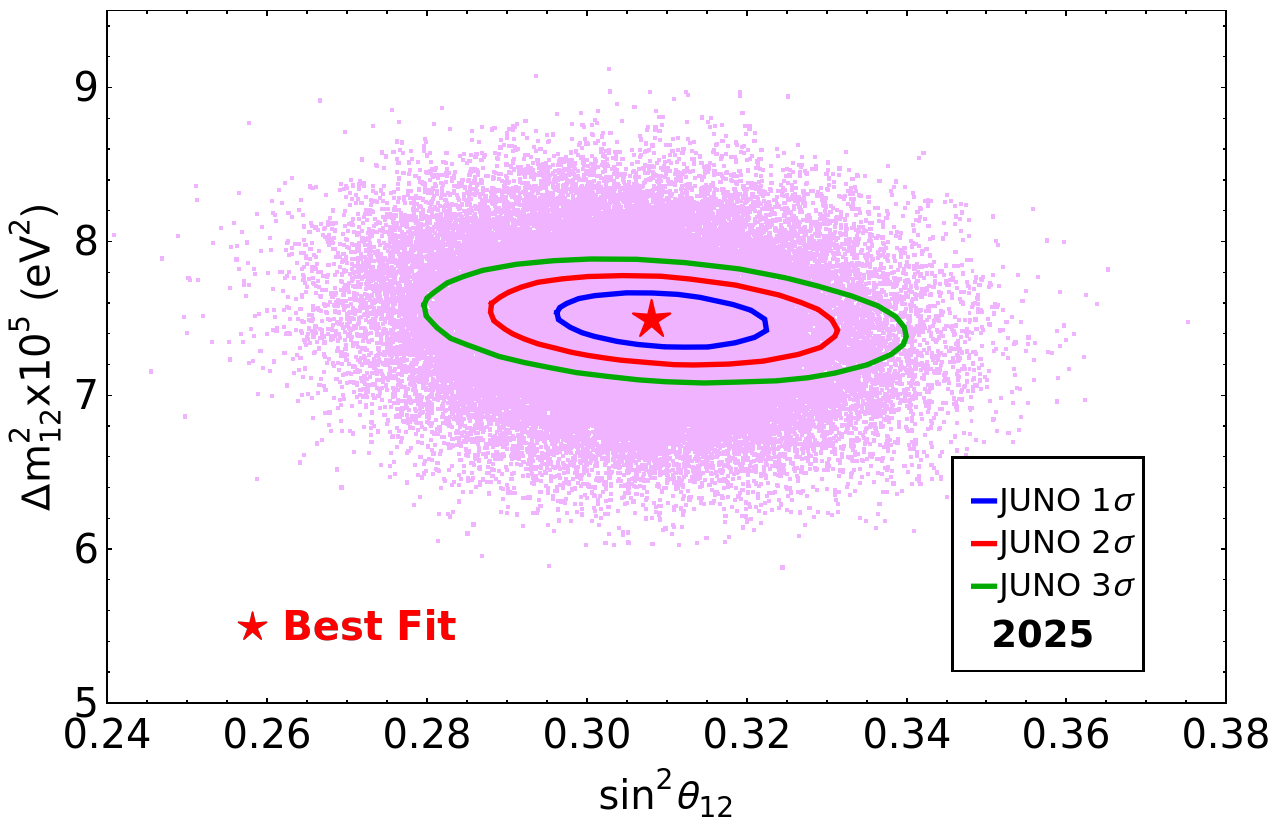}
\includegraphics[width=0.46\textwidth]{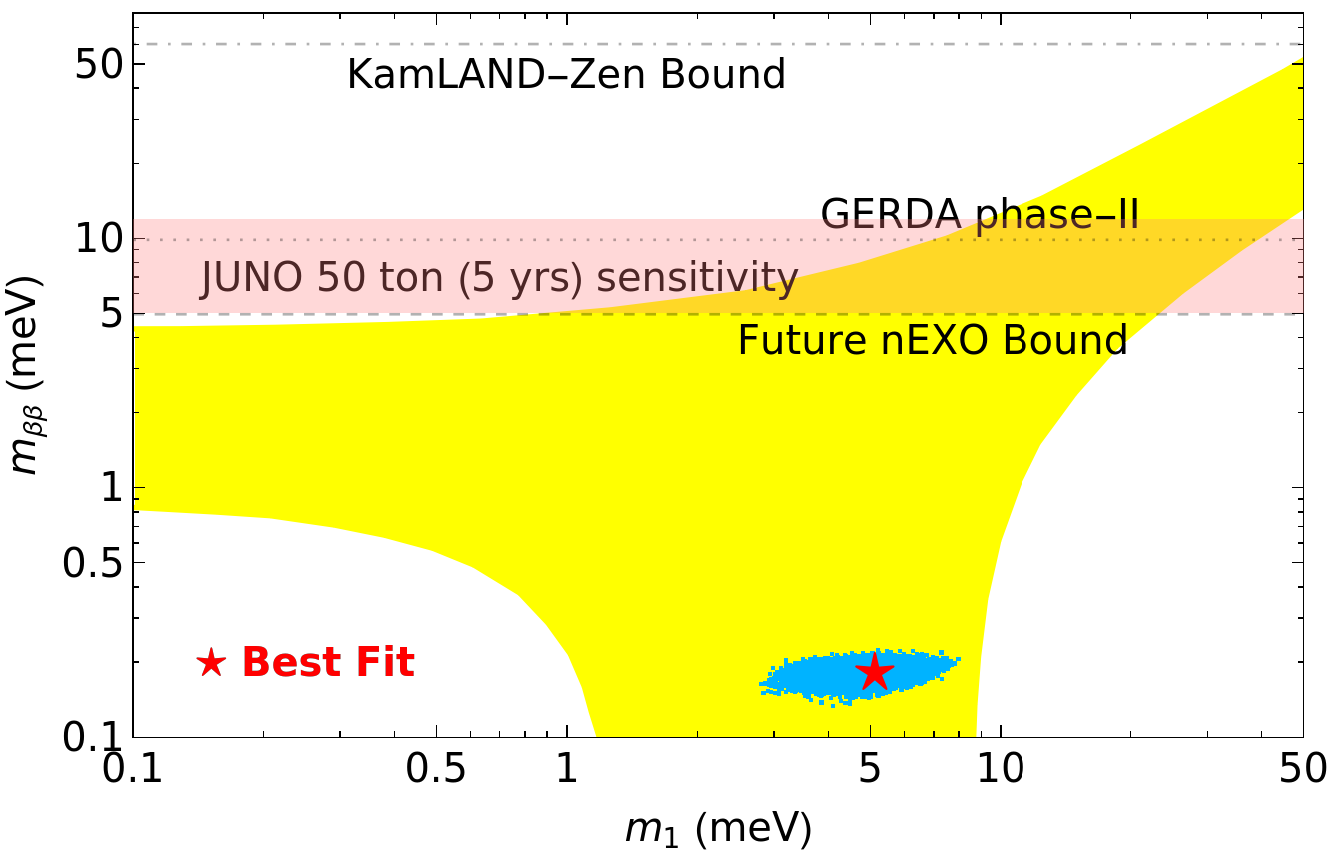}
\caption{MCMC result. Left panel:  Compatibility of our results with the new JUNO result~\cite{JUNO:2025gmd}. Right panel: The likelihood range of neutrinoless double beta decay as a function of the lightest neutrino mass. The current bound from KamLAND-Zen~\cite{KamLAND-Zen:2016pfg} and the future sensitivities of next-generation experiments, such as GERDA Phase II~\cite{GERDA:2019cav} and nEXO~\cite{nEXO:2021ujk}, as well as JUNO's 50-ton sensitivity after 5 years of operation~\cite{Zhao:2016brs}, are shown. Best fit corresponds to $m_1=5.12$ meV and $m_{\beta\beta}=0.18$ meV. } \label{fig:101}
\end{figure}

\begin{figure}[b!]
\centering

\includegraphics[width=0.45\textwidth]{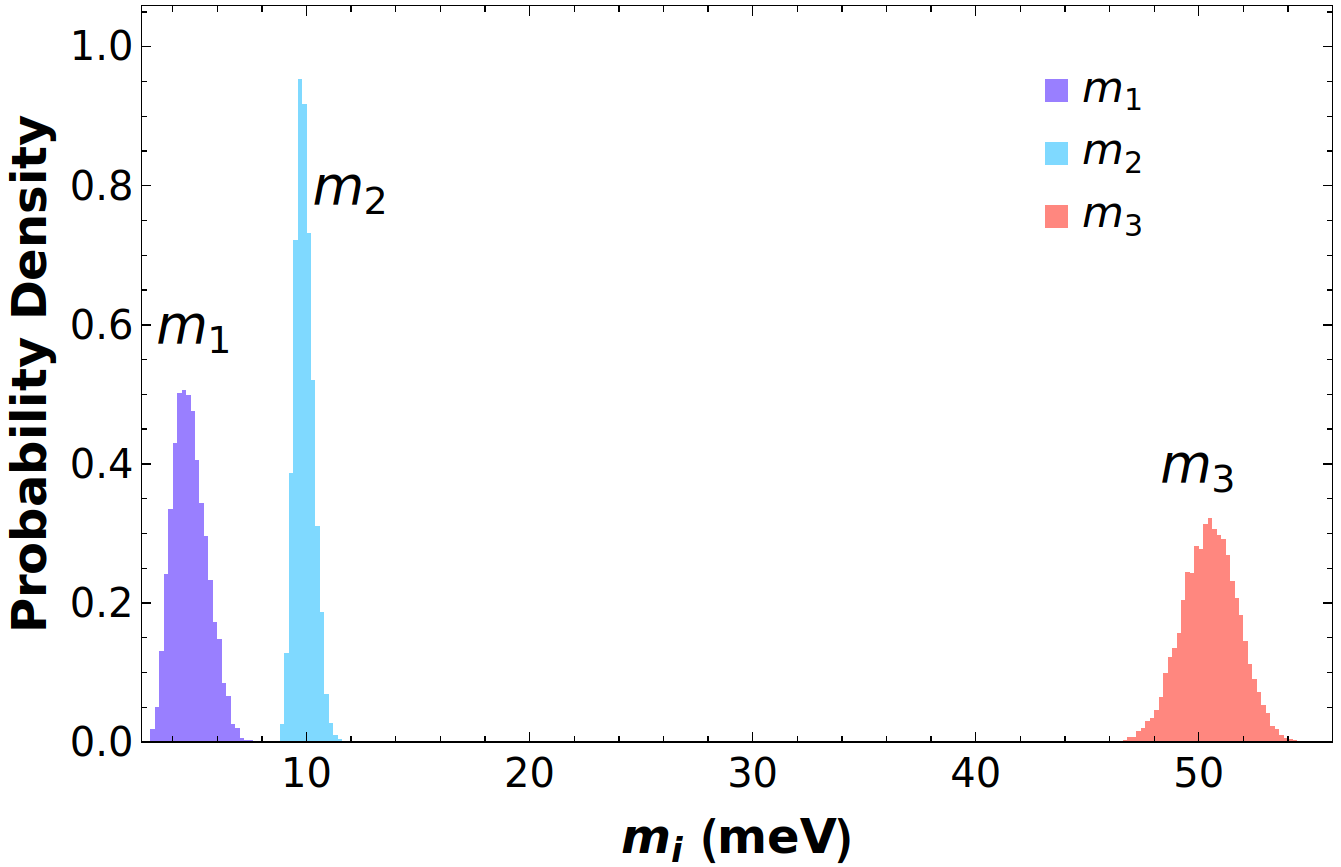}
\includegraphics[width=0.45\textwidth]{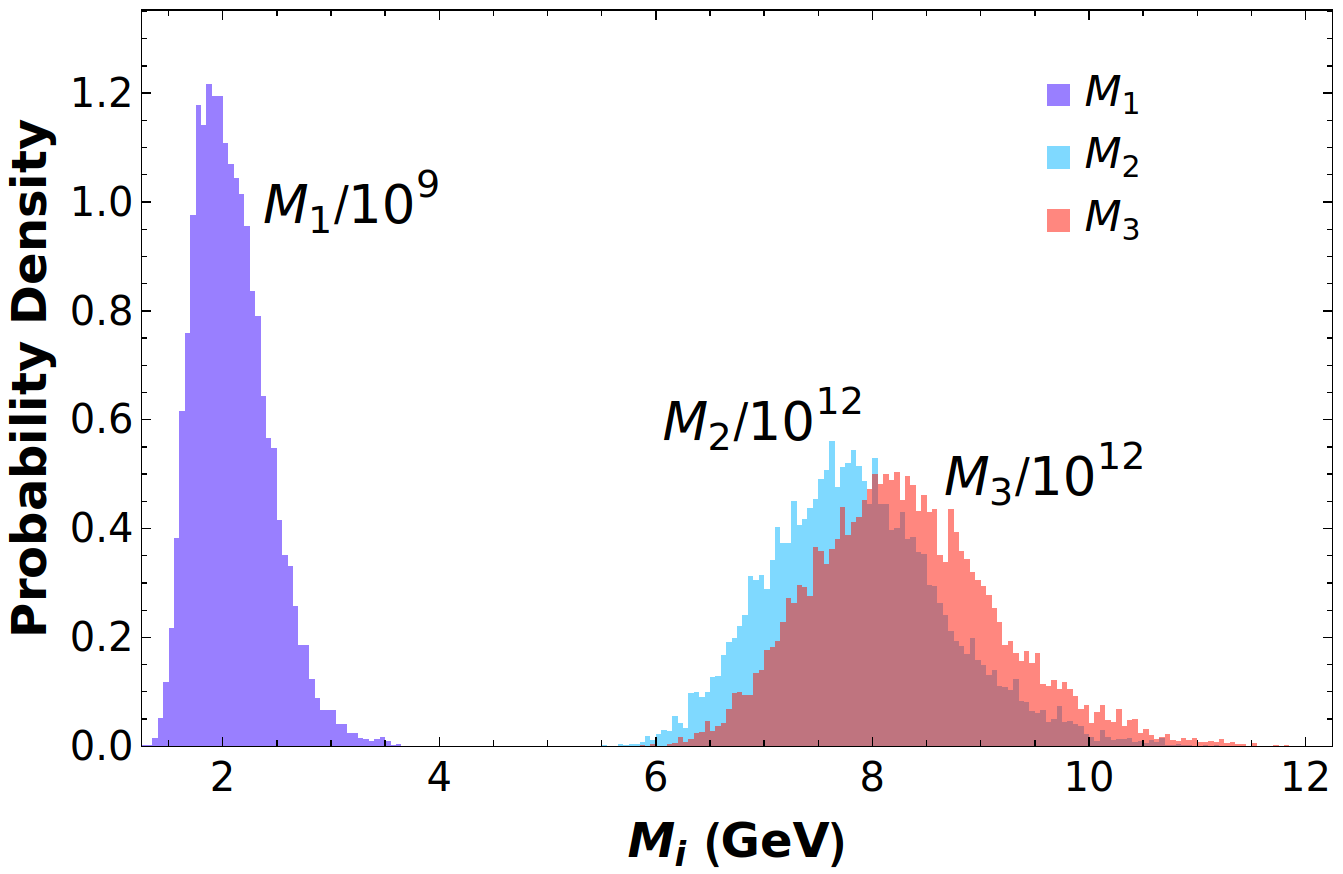}
\caption{MCMC result: The likelihood distribution of light (heavy) neutrino masses in the left (right) panel. The lightest SM neutrino mass is predicted to be in the range  $m_1\sim (2-7)$ meV. In the right panel, the overlapping region with $M_3>M_2$ is indicated by the darker color. } \label{fig:002}
\end{figure}

\begin{figure}[t!]
\centering
\includegraphics[width=0.43\textwidth]{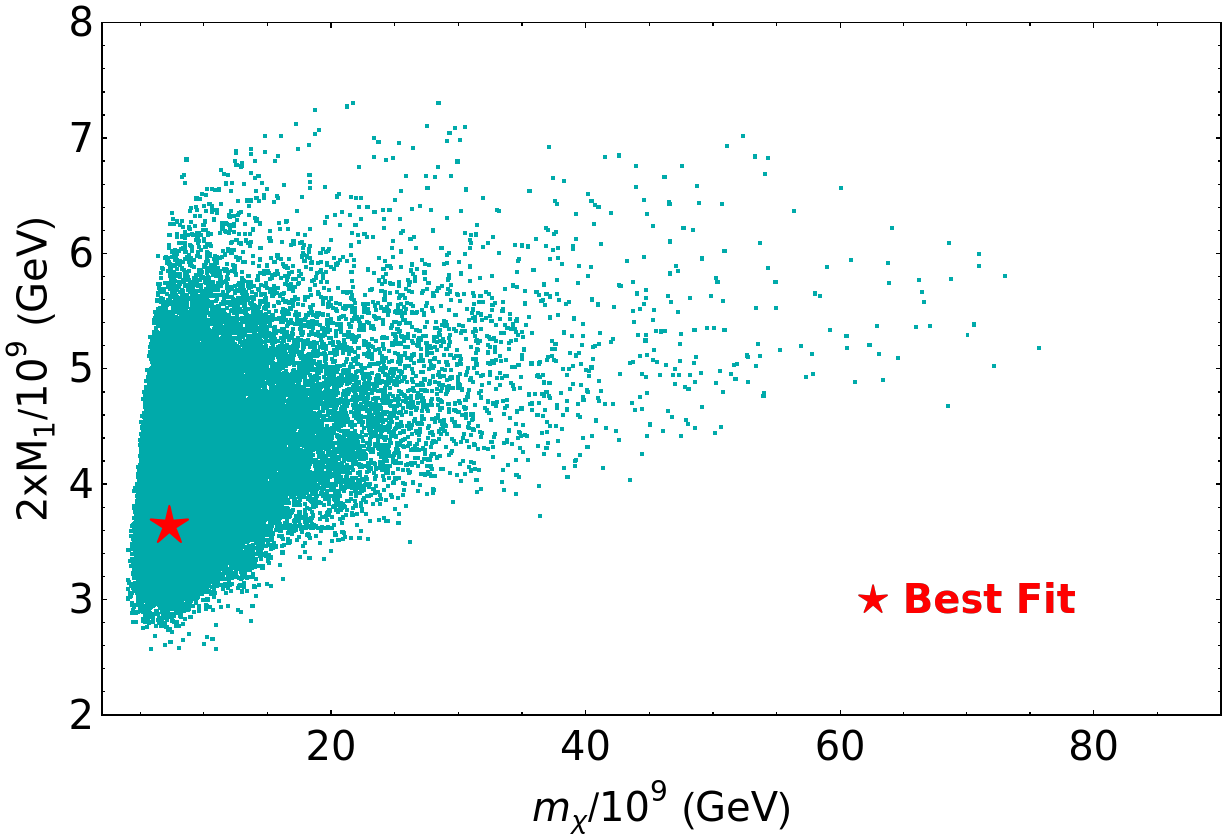}
\caption{MCMC result: likelihood-predicted mass of the inflaton. Best fit corresponds to $m_\chi=7.29\times 10^9$ GeV.} \label{fig:003}
\end{figure}

\begin{figure}[b!]
\centering
\includegraphics[width=0.47\textwidth]{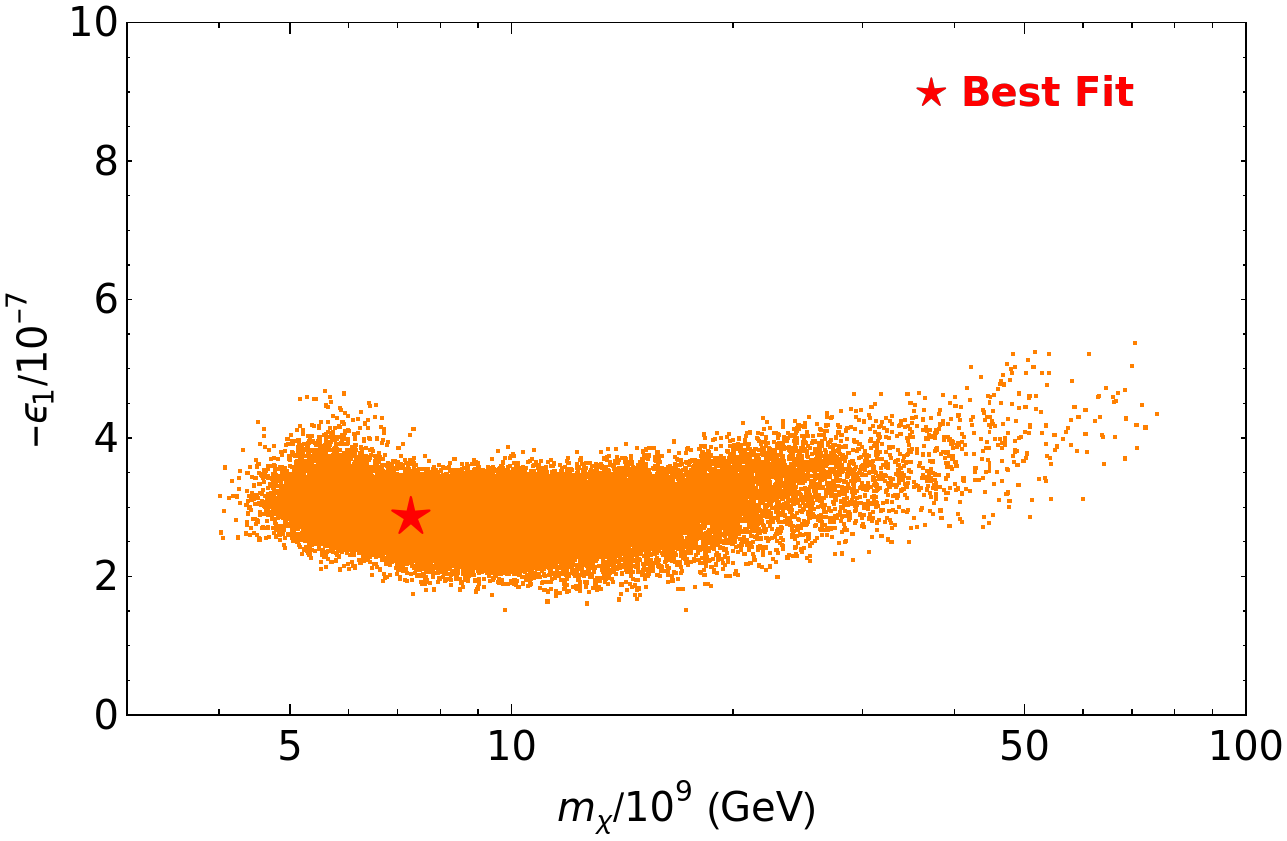}
\includegraphics[width=0.47\textwidth]{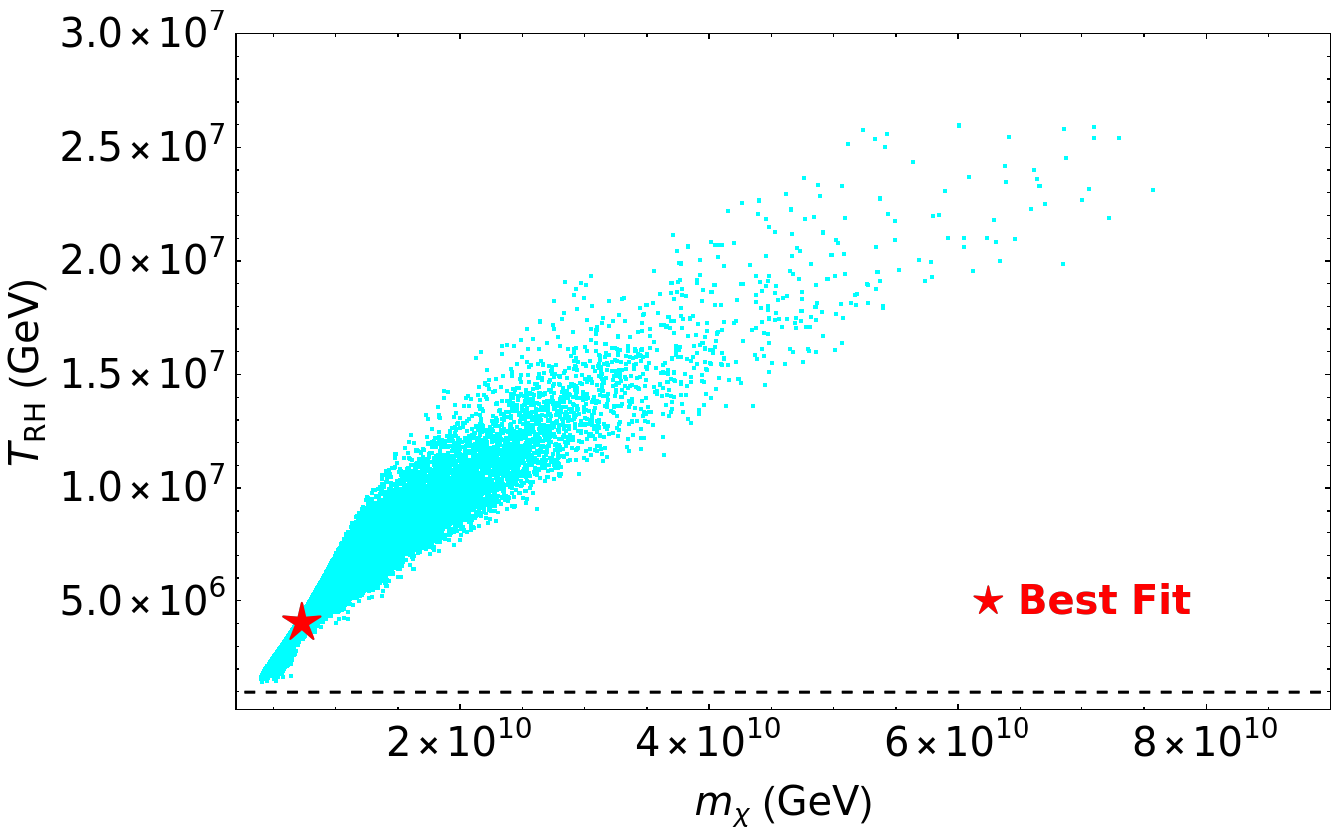}
\caption{MCMC result: The obtained CP-asymmetry parameter (left panel) and the reheating temperature (right panel) as a function of the inflation mass. Best fit solution corresponds to $\epsilon_1=-2.87\times 10^{-7}$ and $T_\mathrm{RH}=4.08\times 10^6$ GeV.    } \label{fig:004}
\end{figure}

\begin{figure}[t!]
\centering
\includegraphics[width=0.46\textwidth]{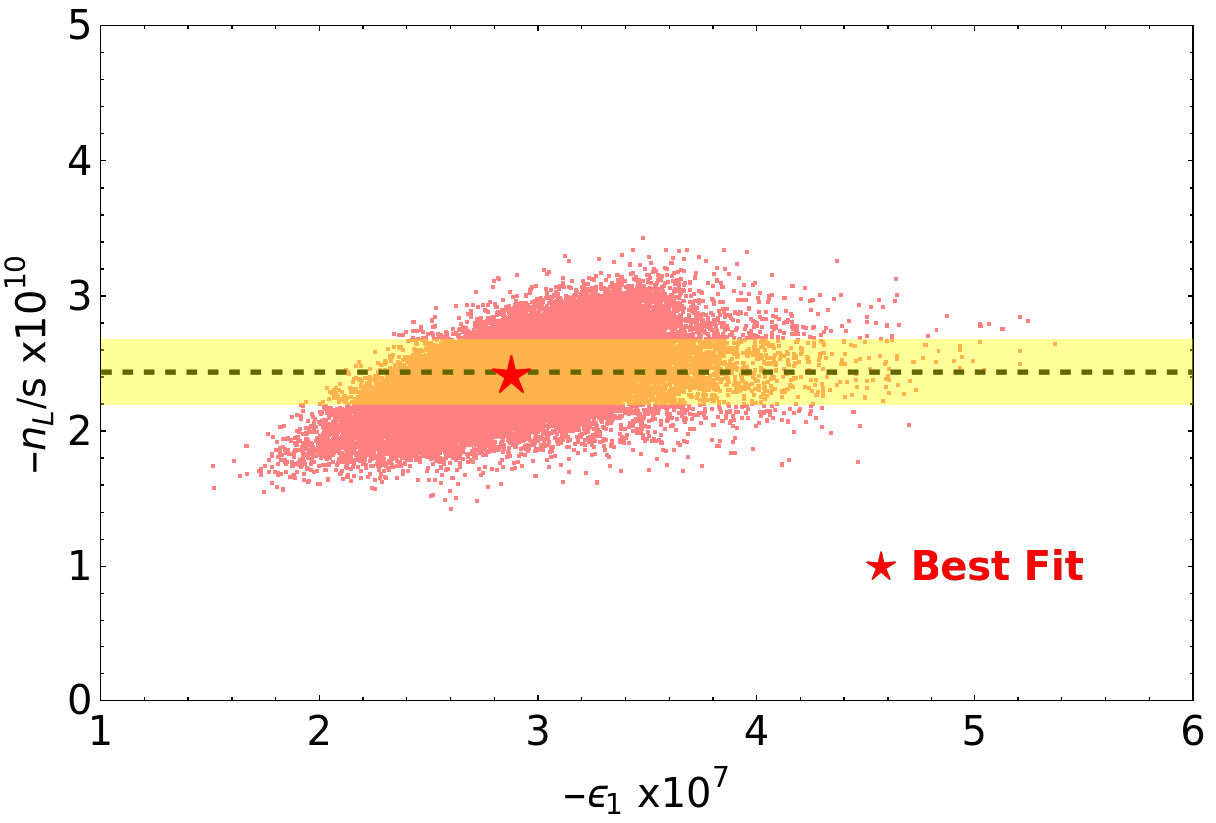}
\caption{MCMC result: The range of the CP-asymmetry parameter compatible with the observed baryon asymmetry. Best fit corresponds to $n_L/s= -2.4\times 10^{-10}$ and $\epsilon_1=-2.87\times 10^{-7}$. } \label{fig:005}
\end{figure}

\begin{figure}[th!]
\centering
\includegraphics[width=0.45\textwidth]{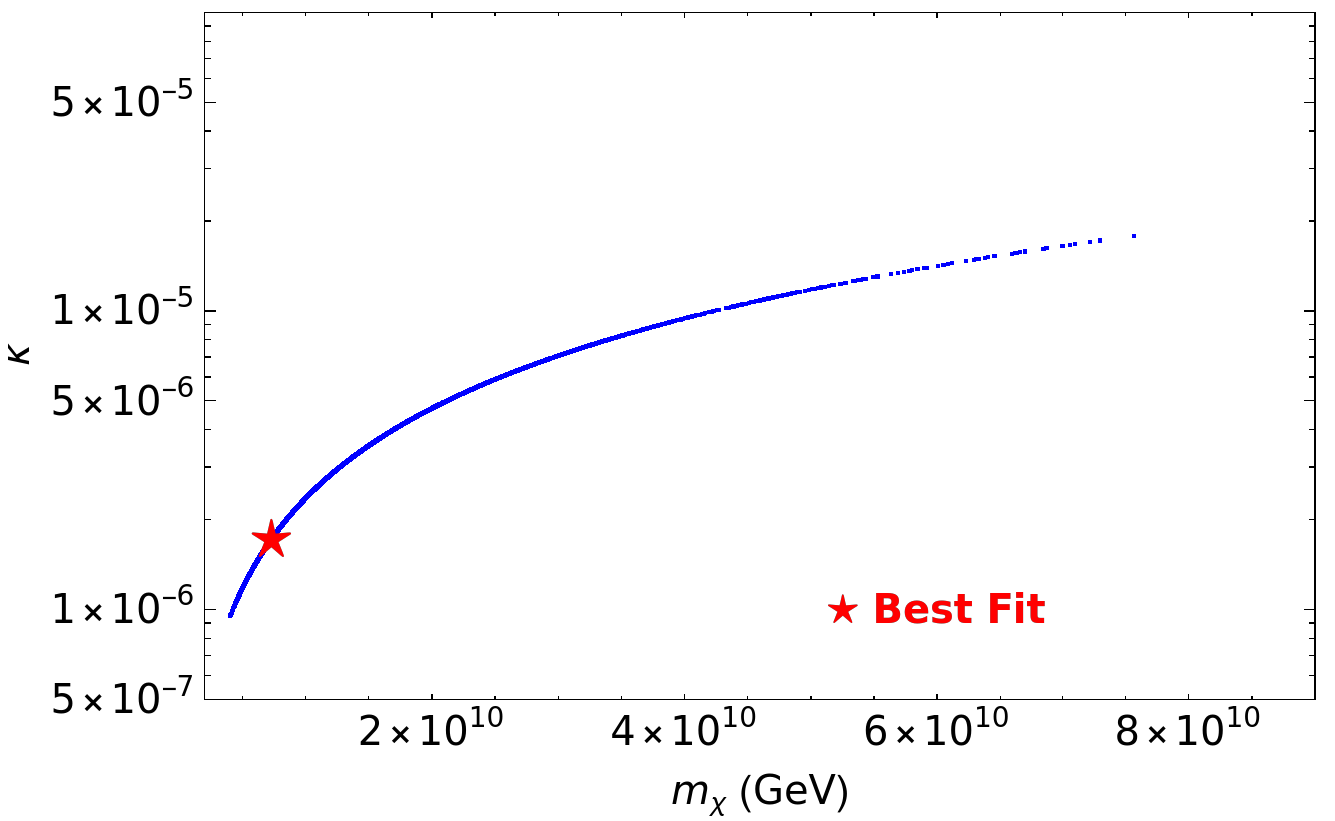}
\caption{MCMC result: The predicted likelihood range of $\kappa= m_\chi/(\sqrt{2} M)$ appearing in the inflationary superpotential Eq.~\eqref{eq:inf}. Best fit corresponds to $m_\chi=7.29\times 10^9$ GeV and $\kappa=1.72\times 10^{-6}$. } \label{fig:006}
\end{figure}

Although Table~\ref{tab:fit} presents the best-fit predictions for some of the physical quantities, a more comprehensive exploration of the parameter space requires investigating the likelihood distributions of all observables in this theory. To this end, once the best-fit solution is obtained, we perform a dedicated Markov Chain Monte Carlo (MCMC) analysis starting from this solution to explore the surrounding parameter space. The results of this MCMC analysis are presented in Figs.~\ref{fig:001}-\ref{fig:006}.

Fig.~\ref{fig:001} shows the likelihood-predicted region for the leptonic CP-violating Dirac phase $\delta_\mathrm{PMNS}$, along with its correlations with the baryon asymmetry parameter $\eta_B$ and the neutrinoless double beta decay parameter $m_{\beta\beta}$. Our model predicts $\delta_\mathrm{PMNS} \sim 100^\circ - 300^\circ$, fully compatible with the current global-fit 1$\sigma$ range (shown as the green vertical band in Fig.~\ref{fig:001}), and $m_{\beta\beta} \sim \mathcal{O}(0.2)\,\mathrm{meV}$, which lies well below the current and future experimental sensitivities. Moreover, the consistency of our results with JUNO’s very recent first measurements in the $\Delta m^2_{12}$–$\sin^2\theta_{12}$ plane is shown in the left panel of Fig.~\ref{fig:101}. The right panel of Fig.~\ref{fig:101} shows the predicted likelihood range of neutrinoless double beta decay as a function of the lightest neutrino mass. In the same figure, together with the current experimental limit on $m_{\beta\beta}$ from KamLAND-Zen~\cite{KamLAND-Zen:2016pfg} (dotted–dashed line), we also present the future sensitivities of next-generation experiments such as GERDA Phase II~\cite{GERDA:2019cav} (dotted line), JUNO~\cite{Zhao:2016brs} (pink shaded region), and nEXO~\cite{nEXO:2021ujk} (dashed line).   Fig.~\ref{fig:002} depicts the distributions of both light and heavy Majorana neutrino masses obtained from the MCMC analysis. This indicates that the lightest SM neutrino mass is likely in the range $m_1 \sim 2 - 7\,\mathrm{meV}$.

The predicted narrow likelihood range of the inflaton mass $m_\chi$, which remains close to $(3-4) M_1$, is shown in Fig.~\ref{fig:003}. In addition, the obtained range of the CP-asymmetry parameter $\epsilon_1$ of Eq.~\eqref{eq:epsilon} and the likelihood-predicted range of the reheating temperature $T_\mathrm{RH}$ as a function of the inflation mass from the MCMC analysis are also presented in Fig.~\ref{fig:004}. Intriguingly, not only is the predicted range of $T_\mathrm{RH}$  compatible with the bound arising from gravitino constraints, but also the expected order,  $T_\mathrm{RH}\sim 10^{6}-10^{7}\,\mathrm{GeV}$, for multi-TeV scale SUSY breaking is recovered. Fig.~\ref{fig:005}  depicts the range of values of the CP-asymmetry parameter compatible with the observed baryon asymmetry. Finally, Fig.~\ref{fig:006} shows the predicted values of $\kappa$ in the $10^{-6} - 10^{-5}$ range, highlighting the significance of non-minimal K\"ahler potential contributions in adjusting the spectral index to be compatible with the experimental data.

\begin{figure}[th!]
\centering
\includegraphics[width=0.45\textwidth]{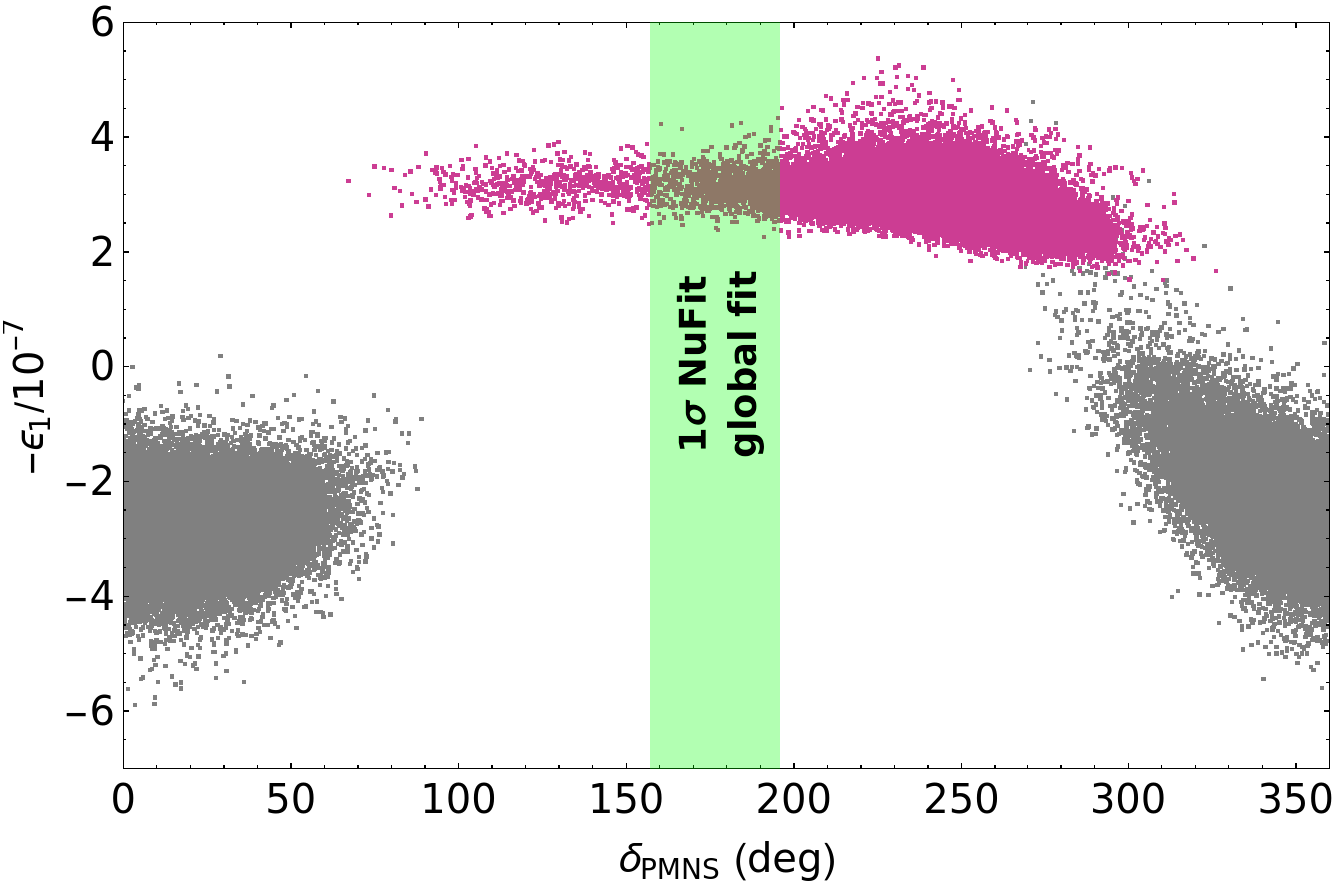}
\caption{  MCMC results with (dark magenta points) and without (gray points) imposing the baryon asymmetry constraints. As evident from these results, the model predicts $|\epsilon_1|_\mathrm{max} \sim \mathcal{O}(10^{-7})$, which is solely determined by the requirements of replicating fermion mass spectrum.} \label{fig:007}
\end{figure}

Finally, we demonstrate that our model predicts $|\epsilon_1|_\mathrm{max} \sim \mathcal{O}(10^{-7})$. To this end, we perform an additional MCMC analysis that is unbiased by the baryon asymmetry constraint and, consequently, independent of the $m\chi$ parameters. In this MCMC, $\epsilon_1$ is allowed to take any value, with both positive and negative signs, as long as the model parameters successfully reproduce the observed fermion masses and mixings. Figure~\ref{fig:007} presents the full results of this analysis, showing the CP-asymmetry parameter, $\epsilon_1$, as a function of the observable  CP-violating phase, $\delta_\mathrm{PMNS}$, in the neutrino sector. Interestingly, the figure reveals that the model accommodates the entire range of $\delta_\mathrm{PMNS} \in (0^\circ$–$360^\circ)$, consistent with the fermion mass fit. The dark magenta points correspond to the subset that reproduces, within 10\%, the correct baryon asymmetry, while the gray points do not. This broad parameter-space exploration makes it clear that $|\epsilon_1|_\mathrm{max} \sim \mathcal{O}(10^{-7})$ is a robust prediction of the model. Moreover, imposing the requirement to reproduce the correct baryon asymmetry with the proper sign further restricts $\delta_\mathrm{PMNS}\in 100^\circ$–$300^\circ$.

Before concluding, we comment on gauge coupling unification within this scenario. The renormalization group evolution of the gauge couplings $\alpha_i = g_i^2/(4\pi)$ at one loop is given by $\alpha_i^{-1}(\mu_{\mathrm{high}}) = \alpha_i^{-1}(\mu_{\mathrm{low}}) + b_i/(2\pi)\,\ln(\mu_{\mathrm{high}}/\mu_{\mathrm{low}})$. Using the experimentally measured values of the SM gauge couplings, one obtains their corresponding values at the SUSY scale, which are~\cite{Antusch:2013jca} $(g_3, g_2, g_1) = (1.0049,\, 0.63436,\, 0.470767)$, for $M_{\mathrm{SUSY}} = 3$ TeV. From the SUSY scale up to the left--right symmetry breaking scale $M_\mathrm{LR} = 3\times 10^{15}$ GeV, the running follows the MSSM RGEs with one-loop coefficients $(b_{Y}, b_{L}, b_{C}) = (33/5,\, 1,\,-3)$. At this scale, the only non-trivial matching condition reads $\alpha_Y^{-1}(M_\mathrm{LR}) = \frac{3}{5}\alpha_{R}^{-1}(M_\mathrm{LR}) + \frac{2}{5}\alpha_{B-L}^{-1}(M_\mathrm{LR})$.  Above $M_{\mathrm{LR}}$ and up to the unification scale $M_{\mathrm{GUT}}$, the one-loop coefficients are $(b_{C}, b_{L}, b_{R}, b_{B-L}) = (-3,\, 1,\, 2,\, 15/2)$, where the MSSM states together with the superfields responsible for breaking the intermediate symmetries are included. Interestingly, we find a relatively precise unification  with the choice $\alpha_{B-L}^{-1}(M_{\mathrm{LR}}) = 28.5$, which leads to $\alpha_{B-L}^{-1}(M_{\mathrm{GUT}}) = 26.2$ and $\alpha_{C}^{-1}(M_{\mathrm{GUT}}) = \alpha_{L}^{-1}(M_{\mathrm{GUT}}) = \alpha_{R}^{-1}(M_{\mathrm{GUT}}) = 26.5$. 

\section{Conclusions}\label{sec:5}
We have explored the phenomenology of a realistic supersymmetric SO(10) model in which higher dimensional operators in the superpotential play an important role in explaining the SM fermion masses, mixings and hierarchies. Extending an earlier model, we require in this work that the observed baryon asymmetry is explained via non-thermal leptogenesis. We briefly discuss how this requirement can be implemented in the framework of supersymmetric hybrid inflation, which allows one to explore in greater detail the leptonic sector of the model. We estimate the masses of the three SM light (with the lightest mass in the range  $m_1\sim (2-7)$ meV) and three right-handed neutrinos, and find that the inflaton mass is about 3-4 times larger than the lightest right handed neutrino mass ($\sim 10^9$ GeV). The reheating temperature $T_\mathrm{RH}$ in this class of inflationary models turns out to be of order $10^6-10^7$ GeV, which is compatible with the well known gravitino constraint on  $T_\mathrm{RH}$ for multi-TeV scale supersymmetry breaking. The best fit value of $\delta_\mathrm{PMNS}\approx 235^\circ$, and for the neutrinoless double beta decay parameter we find $m_{\beta\beta}\approx  0.18$ meV. A Markov Chain Monte Carlo analysis provides a broad range of acceptable $\delta_\mathrm{PMNS}$ values ($100^\circ$--$300^\circ$), but leaves largely intact the prediction of $m_{\beta\beta}$ and the three right-handed neutrino masses. Our statistical analysis, which provides the likelihood-predicted ranges of the observables, agrees fully with JUNO’s recently released first measurement of reactor neutrino oscillations in the $\Delta m^2_{12}$–$\sin^2\theta_{12}$ plane, where JUNO has improved the precision by a factor of 1.6 compared to the combination of all previous measurements.

\subsection*{Acknowledgments}
SS acknowledges the financial support
from the Slovenian Research Agency (research core funding No. P1-0035 and N1-0321). 
\bibliographystyle{style}
\bibliography{reference}
\end{document}